\documentclass[runningheads]{llncs}
\usepackage[utf8]{inputenc}
\usepackage[T1]{fontenc}
\usepackage[english]{babel}
\usepackage{dsfont}
\usepackage{amsmath}
\usepackage{graphicx}
\usepackage{subfigure}
\usepackage{multicol}
\usepackage{multirow}

\begin{document}
\title{More unlabelled data or label more data?\\
A study on semi-supervised\\
laparoscopic image segmentation}

\titlerunning{More unlabelled data or label more data?}
\author{
Yunguan Fu\inst{1,2} \and 
Maria R. Robu\inst{1} \and 
Bongjin Koo\inst{1} \and \\ 
Crispin Schneider\inst{3} \and 
Stijn van Laarhoven\inst{3} \and 
Danail Stoyanov\inst{1} \and \\
Brian Davidson\inst{3} \and 
Matthew J. Clarkson\inst{1} \and 
Yipeng Hu\inst{1}}
\authorrunning{Y. Fu et al.}
\institute{Wellcome/EPSRC Centre for Interventional \& Surgical Sciences and \\
Centre for Medical Image Computing, University College London, London, UK \and
InstaDeep, London, UK \and
Division of Surgery \& Interventional Science, \\
University College London, London, UK
}
\maketitle
\begin{abstract}
Improving a semi-supervised image segmentation task has the option of adding more unlabelled images, labelling the unlabelled images or combining both, as neither image acquisition nor expert labelling can be considered trivial in most clinical applications. With a laparoscopic liver image segmentation application, we investigate the performance impact by altering the quantities of labelled and unlabelled training data, using a semi-supervised segmentation algorithm based on the mean teacher learning paradigm. We first report a significantly higher segmentation accuracy, compared with supervised learning. Interestingly, this comparison reveals that the training strategy adopted in the semi-supervised algorithm is also responsible for this observed improvement, in addition to the added unlabelled data. We then compare different combinations of labelled and unlabelled data set sizes for training semi-supervised segmentation networks, to provide a quantitative example of the practically useful trade-off between the two data planning strategies in this surgical guidance application.
\keywords{Semi-supervised  \and Laparoscopic video \and Image segmentation.}
\end{abstract}

\section{Introduction} \label{sec:introduction}

Deep convolutional neural networks have been proposed to segment livers from surgical video images\cite{gibson2017deep}, a significant step towards fully-automated computer-assisted guidance for liver resection procedures. The automatically segmented liver surfaces can be used to reconstruct anatomical structures for assisting real-time navigation and for registering with preoperative 3D medical images, such as diagnostic CT or MR, to locate the target of operative interest. Precise image-guidance has the potential to increase the number of patients that can be offered laparoscopic liver resection over open surgery, thereby significantly reducing the surgery-related stress and risk.

Further improving the segmentation accuracy may resort to more labelled data or unlabelled data with semi-supervised learning. Like many other medical image segmentation tasks, deep-learning-based approaches often require a substantial amount of labelled data for training, which rely on human experts with specialised clinical knowledge and multidisciplinary experience. On the other hand, acquiring more unlabelled image data from more patients or prolonging procedures may have a significant impact on workflow and patient safety. The data planning decision in relation to performance improvement needs to be weighted by the unit costs associated with these choices, labelling more data and collecting more unlabelled data.

Semi-supervised approaches have been successfully applied in medical image segmentation \cite{bai2017semi,perone2018deep,cheplygina2019not}. However, comparing semi-supervised methods directly with the supervised counterparts has to consider multiple factors, such as added unlabelled data and a different network with its training strategy that is often more complex and specific to application. We postulate that this could lead to inconclusive correlation between confounding factors and the observed performance improvement. 
Based on the `mean teacher' method \cite{mean-teacher}, which has been adapted into several medical imaging applications \cite{perone2018deep,adapted-mean-teacher}, we decomposed the effects into those caused by the change of network (training and architectures) and those by adding unlabelled data. The mean teacher approach averages model weights to produce perturbed predictions as pseudo labels for regularising the training \cite{pseudo-label}, a strategy that can be applied with or without ground-truth labels. In this work, we use the aforementioned surgical application as a real-world example to provide a quantitative analysis of the performance impact on the quantities of labelled and unlabelled training data.

Using real patient data from liver surgery cases, we summaries the contributions in this study as follows: a) A statistically significant higher segmentation accuracy is reported in terms of Dice score and Hausdorff distance, compared with a previously proposed supervised method \cite{gibson2017deep}; b) We demonstrate the possibility that the change of training strategy specific to semi-supervised learning could result in significantly better segmentation results without adding any labelled or unlabelled data; c) We show that adding more unlabelled data potentially can reach the improvement made with more labels, providing a practically important quantitative basis for data planning decisions.

\section{Method} \label{sec:method}
\subsection{Supervised Segmentation Network Architecture} \label{subsec:network-architecture}

\begin{figure}
\centering
\includegraphics[width=\textwidth]{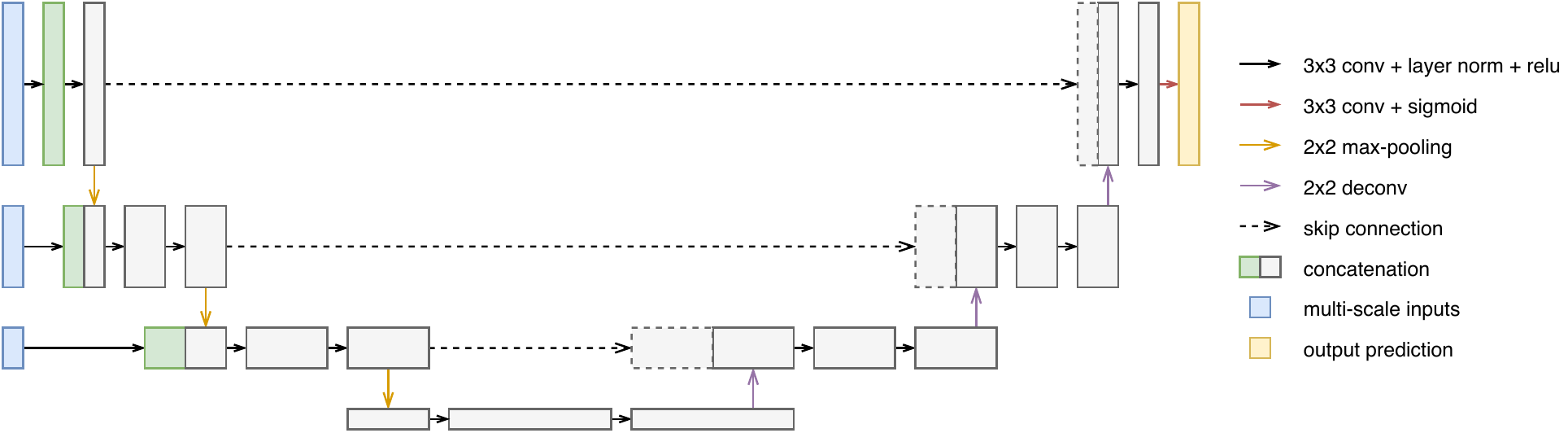}
\caption{U-Net architecture with multi-scale inputs (depth = 3).} \label{fig:u-net}
\end{figure}
\begin{figure}[!ht]
\centering
\includegraphics[width=0.7\textwidth]{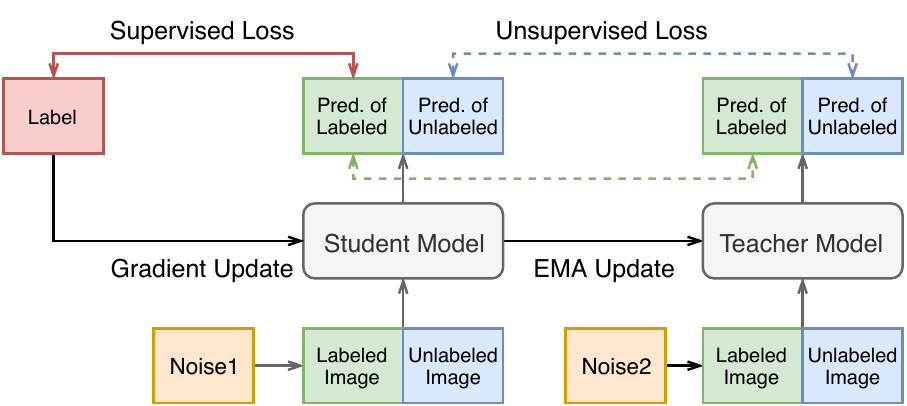}
\caption{Mean Teacher Structure} \label{fig:mean-teacher}
\end{figure}
To analyse the effect with different training data set sizes in this work, we consistently use an exemplar neural network throughout our experiment, which is adapted from a U-Net variant \cite{focal-tversky-loss}.
Like the original U-Net \cite{u-net}, it consists of a downsampling path (encoder) and an upsampling path (decoder), with skip connections added between the two paths. In addition, a multi-scale input image pyramid is added at each encoder layer except for the bottom one. For the decoder, the attention gate and deep supervision are omitted in this network for faster training. The details of the network are illustrated in Fig. \ref{fig:u-net}. The two-class Dice \cite{sudre2017generalised} with $L_2$ regularization is adopted for classifying the foreground pixels representing liver from the background pixels. 

\subsection{Semi-supervised Mean Teacher Training}
Denote the labelled input as $x_l$, with its label as $y_l$, and the unlabelled input as $x_u$. Let $x_{m}=[x_l;x_u]$ be the mixed input. Two identical segmentation networks, the student network $f(x_m,\eta_m^1;\theta_s)$ and the teacher network $f(x_m,\eta_m^2;\theta_t)$ are illustrated in Fig. \ref{fig:mean-teacher}, with different input noise $\eta$ and network weights $\theta$.

During the training, the student network's weights $\theta_s$ are optimized using back-propagated gradients with respect to a regularised segmentation loss:
\begin{align*}
    L_s = L_l(f(x_l,\eta_l^1;\theta_s), y_l) + \lambda~L_u(f(x_m,\eta_m^1;\theta_s), f(x_m,\eta_m^2;\theta_t)),
\end{align*}
where $\lambda$ is a hyper-parameter balancing the contributions of a supervised loss $L_l$ and an unsupervised loss $L_u$, both based on the two-class soft Dice loss \cite{sudre2017generalised}. $L_l$ measures the overlap between the prediction and the ground-truth label, while $L_u$ measures the discrepancy between student and teacher's predictions. The teacher network is updated using exponential moving average (EMA): after each training step, $\theta_t = \alpha \theta_t + (1-\alpha) \theta_s$, where $\alpha$ controls the smoothing. 

One important mechanism of this method is adding noise $\eta_l^i$ and $\eta_u^i$ to labelled and unlabelled image input, respectively, and $\eta_m^i=[\eta_l^i;\eta_u^i]$ for $i\in\{1,2\}$. In this work, we propose to use random affine transformation as the noise in the spatial domain. We apply two independently-drawn affine transformations to the input data as follows: one is applied to the student network input, with the same transformation applied to the available labels for supervised loss; while the second is composed with the first and applied to the teacher network input. The second transformation is then applied to the student network's prediction for computing the unsupervised loss.

\section{Experiment} \label{sec:experiment}
\subsection{Data Set} \label{subsec:dataset}
A total of 41,994 laparoscopic video frames, with a sampling rate of four frames-per-second, were captured from a Storz TIPCAM 3D stereo laparoscope camera in our experiment. These were from thirteen patients during six liver resection and seven liver staging procedures, with informed consents obtained from all patients, and the data collection was approved by our institutional research ethics board. In addition, 2,209 images were selected on which, the regions of liver were manually contoured by an expert clinical research fellow in General Surgery to provide ground-truth segmentation labels. The annotation was performed in NiftyIGI \cite{clarkson2015niftk}, resulting in $67$, $156$, $148$, $168$, $246$, $180$, $140$, $260$, $198$, $178$, $166$, $144$, $158$ labelled frames for each patient respectively.

The original size of frame images were $1920\times540$ pixels in RGB channels with black borders on both sides. For computational and memory efficiency, All images were linearly re-sampled to $128\times384$ for each channel after cropping out the border to a size of $1660\times540$ pixels. 

\subsection{Network Implementation and Training} \label{subsec:details}
The depth of network was $4$ and each network was trained for $10,000$ iterations with a mini-batch size of $32$, using the Adam optimizer with an initial learning rate at $10^{-4}$. The weight of $L_2$ loss was fixed to $10^{-5}$ throughout the experiments. The network output has the same size as the re-sampled input image, larger than $81\times21$ used in previous work \cite{gibson2017deep}. In the loss used in the mean teacher training, $\lambda=0.1\beta$ with $\beta$ increasing progressively, i.e. $\beta=\exp(-5(\max(1-\frac{S}{L},0))^2)$, where $S$ is the current training step and $L=1000$ is the ramp-up length. The EMA decay $\alpha$ was fixed to $0.99$ during the initial ramp-up phase and $0.999$ afterwards. All networks were implemented in TensorFlow and trained using Nvidia Tesla V100 general-purpose graphics process units on a DGX-1 workstation. To avoid over-fitting the entire data set, all the reported hyper-parameter values were configured empirically without extensive tuning.

\subsection{Evaluation} \label{subsec:evaluation}
All experiment results reported in this paper were based on 13-fold leave-one-patient-out cross-validations: for each fold, data from one patient was used for evaluation and the network was trained on the remaining data. The predicted binary masks representing segmentation were first re-sampled to $1660\times540$ and then processed by filling the holes before evaluation. Commonly-adopted data augmentation strategies for surgical video applications, including contrast and brightness adjustment and standardization, were also used before feeding the input data. The segmentation performance was measured by the Dice score and the 95th-percentile Hausdorff distance. The reported Hausdorff distance is in pixels and $100$ pixels correspond approximately $1.5$ mm to $6.0$ mm, depending on the typical object-to-camera distance range in this application.

To test different data set sizes, $2\%$, $10\%$, $25\%$, $50\%$ and $100\%$ of the labelled data set were randomly sampled from each patients for semi-supervised networks, while $0\%$, $6.25\%$, $25\%$ and $100\%$ of the unlabelled data set were sampled with $0\%$ indicating the mean teacher models trained without unlabelled data. A single network without the mean teacher model (hereafter referred to as the \textit{baseline supervised network}\footnote{The mean teacher model without unlabelled images is also fully-supervised.}) was also tested. In practice, however, the availability of the labelled and unlabelled image data would be influenced by other practical factors, such as cost and patient cohort sampling, and is highly application-dependent. This controlled experiment was designed with a simplified condition that excludes potential anatomical-variation-introduced difference between patients and should be considered as the first step towards a more comprehensive experiment design considering both inter- and intra-patient variation. We also report the statistical significance in the observed differences throughout the presented experiments using non-parametric Wilcoxon signed-rank tests at a significance level of 0.05.

\section{Result} \label{subsec:result}

\subsubsection{Baseline Supervised Network (SL)}
The median Dice scores on 13 folds from the baseline supervised network trained using all labelled images ranged from $0.85$ to $0.98$ with a median of $0.95$, compared with $0.78$, $0.98$ and $0.97$ from the previous study \cite{gibson2017deep}, respectively. The difference was probably due to the change of loss function and the adoption of the U-net variant. When varying the quantity of the training (labelled) data from $2\%$ to $100\%$, the segmentation performance was improved, from $0.9250$ to $0.9594$ and from $137.00$ to $91.61$, for Dice score and Hausdorff distance, respectively.

\subsubsection{Mean Teacher (MT)}
\begin{figure}[!ht]
    \centering
    \includegraphics[width=0.49\textwidth]{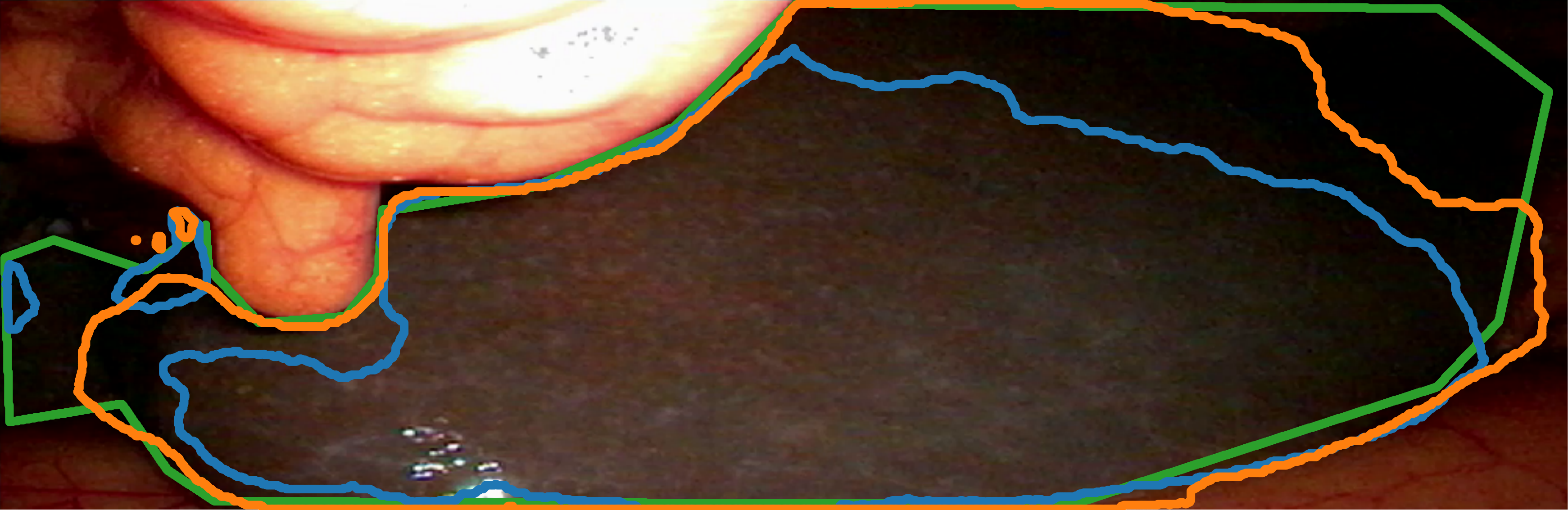}
    \includegraphics[width=0.49\textwidth]{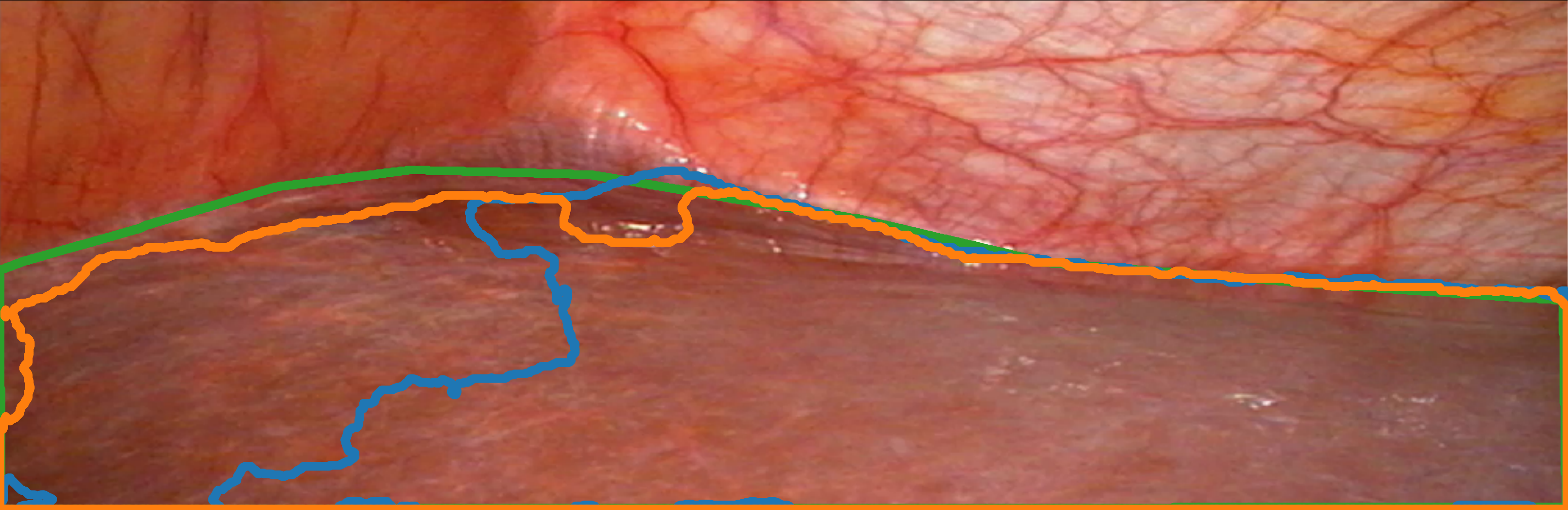}
    \caption{Two examples with ground truth (green) and the predictions of the supervised model (blue) and mean teacher trained with all labelled and unlabelled data (orange).}
    \label{fig:demo}
\end{figure}

The results for SL and MT with $100\%$ unlabelled data are summarised in Table \ref{tab:sl-mt-100}. Both the medians of Dice score and Hausdorff distance from MT were significantly better (both p-values $<0.001$). The median Dice scores on $13$ folds ranged from $0.87$ to $0.98$, with a median of $0.97$, therefore surpassed the previous study \cite{gibson2017deep} (p-value $=0.008$). Examples are shown in Fig. \ref{fig:demo}.

\begin{table}[!ht]
\centering
\caption{Comparison between supervised model (SL) and mean teacher (MT) with all labelled and unlabelled data.}
\label{tab:sl-mt-100}
\begin{tabular}{|c|c|c|c|c|c|}
\hline
~Metric~ & ~Method~ & ~Median~ & ~Mean~ & ~Std~ & Wilcoxon \\ \hline
 \multirow{2}{*}{~Dice Score~} & SL & ~0.9594~ & ~0.8792~ & ~0.1819~ & ~\multirow{2}{*}{9.27e-28}~\\ \cline{2-5}
 & MT & ~\textbf{0.9646}~ & ~\textbf{0.9032}~ & ~0.1483~ & \\ \hline
\multirow{2}{*}{~Hausdorff Distance~} & SL & ~91.61~ & ~148.23~ & ~166.86~ & ~\multirow{2}{*}{4.40e-07}~\\ \cline{2-5}
 & MT & ~\textbf{81.49}~ & ~\textbf{137.57}~ & ~163.16~ & \\ \hline
\end{tabular}
\end{table}
\begin{figure}[!ht]
    \centering
    \includegraphics[width=0.49\textwidth]{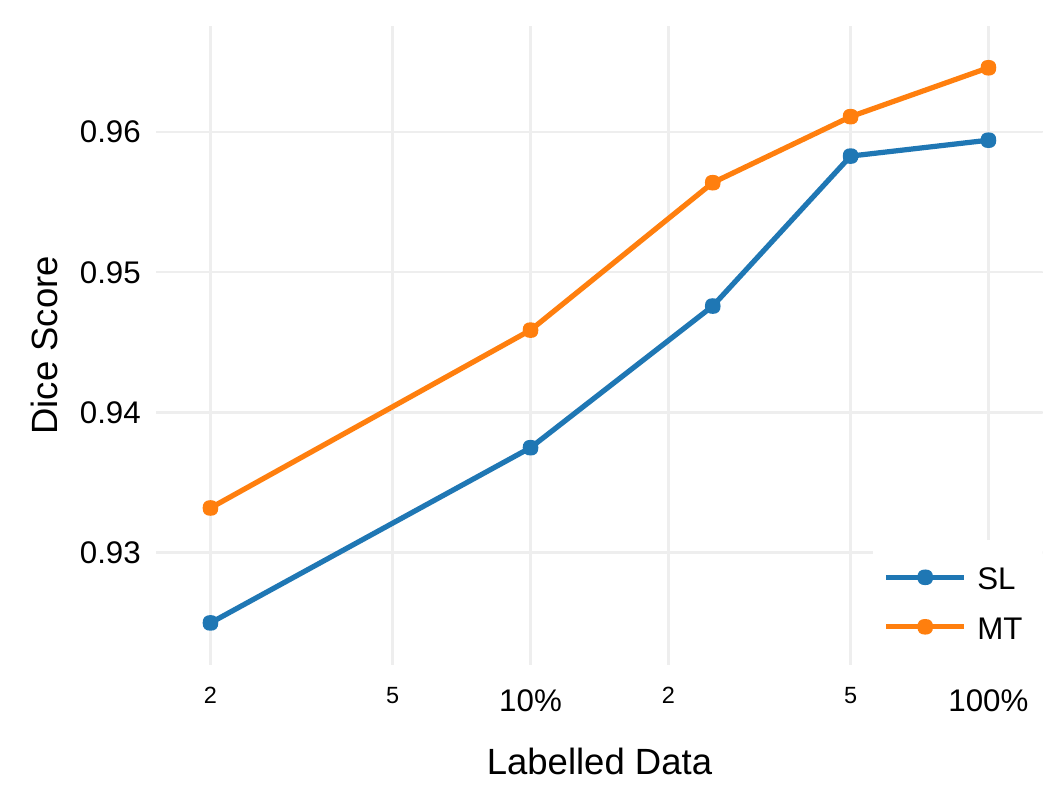}
    \includegraphics[width=0.49\textwidth]{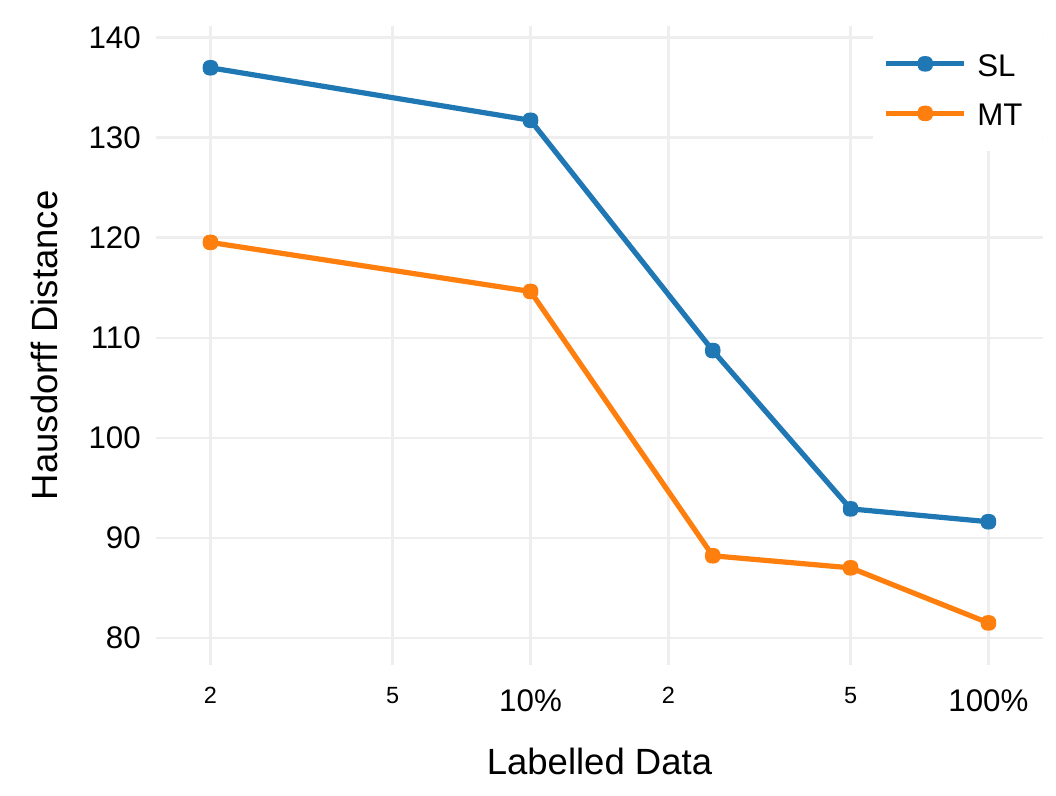}
    \caption{Comparison between supervised model and mean teacher with different quantities of labels.}
    \label{fig:lbl}
\end{figure}

\subsubsection{Mean Teacher with Different Labelled Data Set Sizes}
The median Dice scores for the MT models, trained with all available unlabelled data and different quantities of labelled data, varied from $0.9332$ to $0.9646$. It consistently outperformed SL with the same labelled data set sizes sampled, as shown in Fig. \ref{fig:lbl}. The Hausdorff distance results also showed a consistent difference. In addition, a clear overall trend for both segmentation metrics can be observed: the performance improves as the number of labelled data increases. 

\subsubsection{Mean Teacher with Different Unlabelled Data Set Sizes} 
\begin{figure}[!ht]
    \centering
    \includegraphics[width=0.49\textwidth]{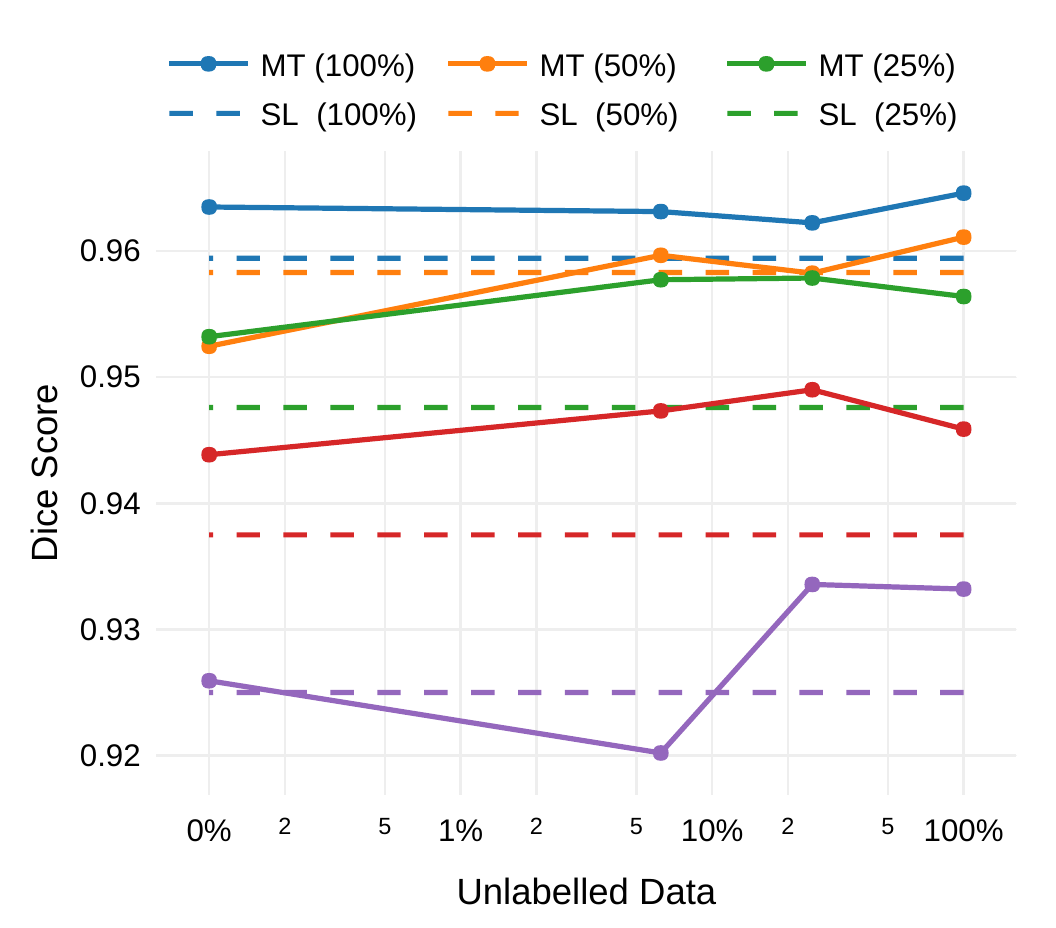}
    \includegraphics[width=0.49\textwidth]{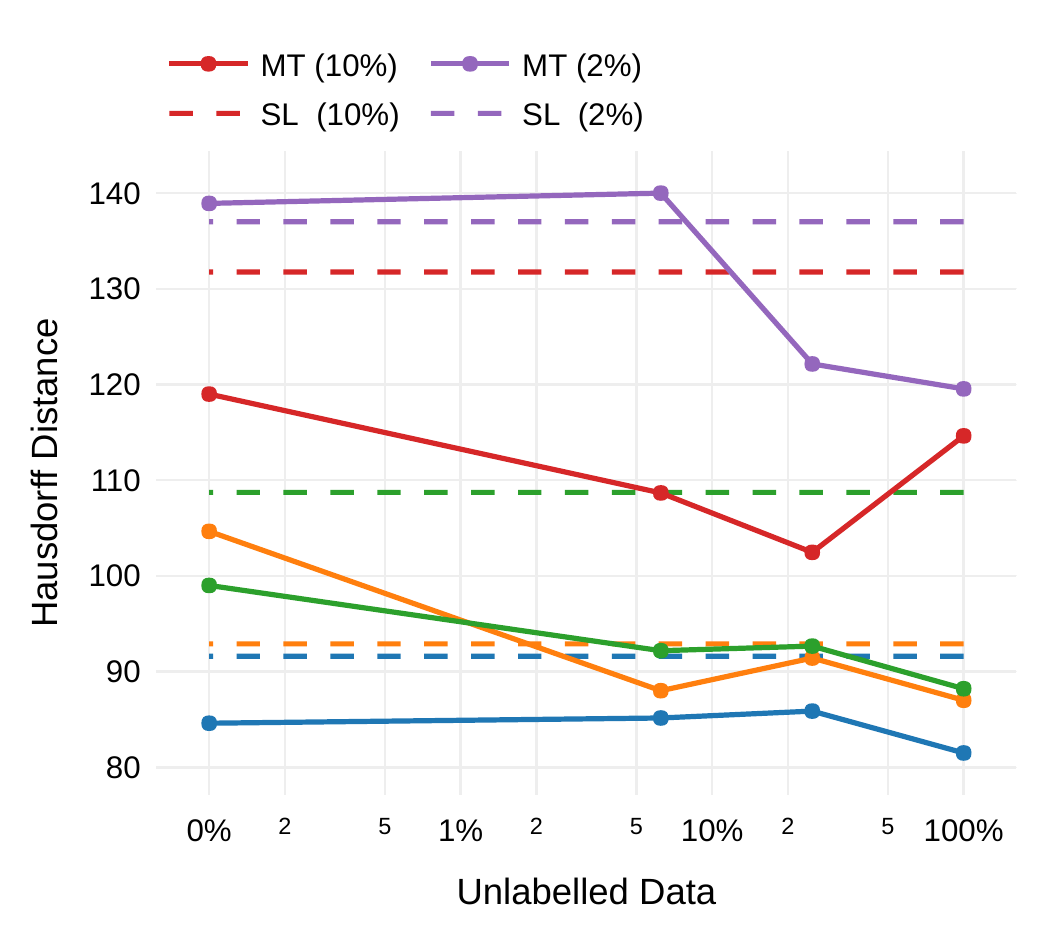}
    \caption{Comparison between supervised model (SL) and mean teacher (MT) with different quantities of unlabelled data. The quantity of labelled data used is indicated in brackets.}
    \label{fig:unlbl}
\end{figure}

Median Dice scores are plotted in Fig. \ref{fig:unlbl} with the quantity of labelled data indicated in the brackets. Without using any unlabelled data, MT generally outperformed SL; with more unlabelled data, MT produced better segmentation in general, but it was not monotonic. For instance, using $6.25\%$ of unlabelled data improves MT ($10\%$) from $0.9438$ to $0.9473$ in terms of Dice score, but for MT ($2\%$) the score decreases from $0.9259$ to $0.9202$. This may be caused by a) high correlation between unlabelled data due to the nature of the procedure and the omitted inter-patient variation (also discussed in Sec. \ref{subsec:evaluation}); b) the lack of optimised semi-supervised training and hyper-parameter tuning, which was not pursued further for the purpose of this work.
Practically important, perhaps more interesting, results can be found to quantify the trade-off between the labelled and unlabelled data. For example, using $100\%$ unlabelled data, MT ($50\%$) reached a Dice score of $0.9611$ which was higher than SL ($100\%$), $0.9594$, depicting a scenario in which more unlabelled data achieve a comparable performance as adding labels would.

\section{Conclusion} \label{sec:conclusion}
The quantified differences showed in this work, such as the improvement due to more labelled and/or unlabelled data, are useful in developing machine learning applications that in turn assist clinical procedures. To summarise, we have shown a statistically significant improvement in segmenting liver from laparoscopic video images using a semi-supervised mean teacher method. Whilst adding more labelled data generally improves the segmentation, it is possible to use more unlabelled data, instead of labelling more data, to achieve comparable level of segmentation accuracy. To the best of our knowledge, it is the first time these conclusions are presented with quantitative evidence based on real patient data.

These results, however, should be interpreted with the limitations of the experiment design due to practical constraints. We suspect that non-optimised semi-supervised training and sampling intra-patient variation, also discussed in Sec. \ref{subsec:details} and \ref{subsec:evaluation}, respectively, are possible reasons for the perturbing segmentation performance as unlabelled data increase, which limited potentially larger improvement. Nevertheless, the reported high segmentation accuracy warrants a high applicability of these presented models for clinical use. Thus, the statistical significance found in the performance changes, measured on independent test data, suggest potential clinical value in planning data for training these semi-supervised models. These experiments also produced a set of quantitative results, on which future work can build on to answer further multidisciplinary questions.

\section*{Acknowledgement}
This work is supported by the Wellcome/EPSRC Centre for Interventional and Surgical Sciences (WEISS) (203145Z/16/Z). DS receives funding from EPSRC [EP/P012841/1]. MC receives funding from EPSRC [EP/P034454/1]. BD was supported by the NIHR Biomedical Research Centre at University College London Hospitals NHS Foundations Trust and University College London. The imaging data used for this work were obtained with funding from the Health Innovation Challenge Fund [HICF-T4-317], a parallel funding partnership between the Wellcome Trust and the Department of Health. The views expressed in this publication are those of the author(s) and not necessarily those of the Wellcome Trust or the Department of Health.
\bibliographystyle{splncs04}
\bibliography{ref}

\end{document}